\documentclass[conference,a4paper]{IEEEtran}
\IEEEoverridecommandlockouts
\usepackage{cite}
\usepackage{amsmath,amssymb,amsfonts}
\usepackage{algorithmic}
\usepackage{graphicx}
\usepackage{textcomp}
\usepackage{xcolor}
\usepackage{url}
\usepackage{arydshln}
\usepackage{float}
\def\BibTeX{{\rm B\kern-.05em{\sc i\kern-.025em b}\kern-.08em
    T\kern-.1667em\lower.7ex\hbox{E}\kern-.125emX}}

\begin{document}

\title{System Integrity Protection Schemes \\in the Nordics -- a comparative analysis
\thanks{This work has been carried out within the framework of the SPS4SE project, which is funded by the Swedish Energy Agency and coordinated by RISE Research Institutes of Sweden AB.}
}

\author{\IEEEauthorblockN{Gabriel Malmer\\Olof Samuelsson}
\IEEEauthorblockA{\textit{Industrial Electrical Engineering and Automation} \\
\textit{Lund University}\\
Lund, Sweden \\
gabriel.malmer@iea.lth.se}
\and
\IEEEauthorblockN{Arvid Rolander\\Lars Nordström}
\IEEEauthorblockA{\textit{Electric Power and Energy Systems} \\
\textit{KTH Royal Institute of Technology}\\
Stockholm, Sweden \\
arvidro@kth.se}
\and
\IEEEauthorblockN{Emil Hillberg\\Susanne Ackeby}
\IEEEauthorblockA{\textit{Electric Power Systems} \\
\textit{RISE Research Institutes of Sweden}\\
Göteborg, Sweden \\
emil.hillberg@ri.se}
}

\maketitle
\IEEEpubid{\begin{minipage}[t]{\textwidth}\ \\[12pt]
    \centering 979-8-3503-9042-1/24/\$31.00~\copyright2024 IEEE
    \end{minipage}}

\begin{abstract}
To increase the utilisation rate of the power system and accelerate electrification while providing a high degree of security and reliability, System Integrity Protection Schemes (SIPS) are of great importance. SIPS functions are automatic remedial actions, detecting abnormal conditions or contingencies in the system and taking control action to mitigate these conditions. Design, implementation, maintenance and coordination of SIPS are all important aspects for desired operation. However, different actors have chosen different approaches to using SIPS for capacity enhancement, and there are discrepancies in how capacity is valued in relation to for example complexity, reliability and risk. Additionally, definitions often vary between countries. This paper reports on a joint survey and interview study on SIPS with stakeholders and experts in the Nordic countries -- including TSOs, DSOs and industry. Combined with a literature review, a comparison and analysis of how SIPS are used in the Nordics is performed, particularly in relation to ENTSO-E capacity allocation.
\end{abstract}


\section{Introduction}
The world is facing an energy trilemma -- namely to provide everyone with reliable, affordable and sustainable energy. One of the most scalable and promising solutions to this trilemma is electrification. To achieve sustainable electrification however, a rapid expansion of both grid capacity and fossil-free electricity generation is needed. Grid capacity has emerged as one of the main challenges to electrification, due to the large costs and times involved in strengthening the grid when production and consumption patterns are changing \cite{IEAelectrification2024}. This has sparked interest in alternatives to traditional grid reinforcement, such as grid-enhancing technologies and system automation \cite{gridliftoff2024}. System Integrity Protection Schemes (SIPS) have an important role to play in this transition, particularly in relation to a growing share of inverter-based resources (IBRs).

However, we perceive that the purpose of SIPS often varies, that the implementation could be standardised and coordinated to a higher degree, and that there is often room for additional deployment of SIPS. In the Nordic countries, which is the region in focus in this report, different actors have chosen different approaches to using SIPS for capacity enhancement, and there are discrepancies in how much capacity is valued in relation to for example complexity, reliability and risk.

\subsection{Aim of study \label{sec:aimofstudy}}
In this paper, we compare three different aspects of SIPS usage in the Nordics, namely purpose; implementation and maintenance; and philosophy and risk assessment. Particularly, we aim to assess and compare how SIPS are used in relation to capacity allocation today. This is done through a literature review, combined with a joint survey and interview study with the key stakeholders and experts in the Nordic countries -- including TSOs, DSOs and industry.

\subsection{Terminology and definitions}

Across the world, in Europe and even within the Nordics, there are different naming conventions
for power system automatics used to prevent disturbance propagation, thus improving operational security under stressed conditions. \textit{System Integrity Protection Schemes} (SIPS), \textit{System Protection Schemes} (SPS), \textit{Special Protection Schemes} (SPS) and \textit{Remedial Action Schemes} (RAS) are all examples of frequently used terms \cite{Stankovic2022}. These different conventions can hinder discussion and knowledge sharing on the topic, hence there is a need to clarify the terminology and definitions.

Previous work on standardising the language regarding SIPS has been made in \cite{Stankovic2022}. We find these naming conventions sound and intend to use them as much as possible. Other useful references include the Cigré report from 2001 on System Protection Schemes in Power Networks \cite{Acker2001}, the IEEE Guide for Engineering, Implementation, and Management of System Integrity Protection Schemes \cite{IEEEguide2020}, and the Cigré proposed framework for coordinated power system stability control \cite{CigreControl2018}.

In our work, we intend to consistently use the term \textit{System Integrity Protection Schemes} (SIPS), due to its uniqueness and adequacy. Another frequently used term is \textit{remedial action} (RA), which in EU grid code (article 2(13), CACM) \cite{EU_CACM} is defined as ``any measure applied by a TSO or several TSOs, manually or automatically, in order to maintain operational security''. We perceive SIPS to be a subcategory of RAs, namely RAs that are curative and automatic. \textit{Mitigative action} is in principle synonymous with RA, but used more in the SIPS context, to specify the output of SIPS.

According to \cite{IEEEguide2020}, SIPS is an umbrella term including SPS and RAS as well as underfrequency, undervoltage and out-of-step protection schemes. We intend to use SIPS in the same broader sense, including e.g. under-frequency load shedding (UFLS). As in \cite{Acker2001}, we also perceive SIPS to consist of some input (critical system condition), some output (mitigative action) and some logic in between. 

As stated in \cite{Acker2001}, SIPS can be used for increased reliability, increased transfer capacity or a combination of the two. In our work, increased transfer capacity is in focus. SIPS for this purpose can be either intermediate solutions, providing acceptable operational security until the grid has been reinforced, or permanent. Another SIPS characteristic is the spatial range: SIPS can either be implemented locally, in a single substation, or be distributed across wide areas. Wide Area Monitoring, Protection And Control (WAMPAC) is for example a form of SIPS \cite{WAMPAC2019}. There is also a difference between event-based SIPS, triggered by discrete conditions such as breaker trips, and response-based SIPS, triggered by measured electric variables such as frequency or voltage \cite{Stankovic2022}.

Unacceptable system conditions that SIPS are designed to mitigate are typically rotor angle instability (small-signal and transient), frequency instability (over- and underfrequency), voltage instability (small and large disturbance), abnormal voltage (over- and undervoltage, deviation) and thermal overload \cite{IEEEguide2020}. Mitigative actions include (but are not limited to) load shedding, generation rejection, generation rescheduling/active power control, var rescheduling/reactive power control, HVDC control, FACTS control and grid reconfiguration.

\subsection{Overview of the Nordic power system}
The Nordic power system is herein defined as the electric power generation, transmission and consumption facilities in the Nordic synchronous area (Sweden, Norway, Finland and Eastern Denmark). However, when talking about the power systems in the Nordic countries, Iceland and Western Denmark are also included. In this work, all TSOs in the Nordic countries have been involved as stakeholders, yet the main focus is on the Nordic synchronous area. The Nordic region has a high electricity intensity, with a per capita consumption more than twice as high as the EU average \cite{IeaEnergyStatistics, worldbank_capitaconsumption}.

The cooperation between the Nordic TSOs has a long history, to a large extent through the cooperative body \textit{Nordel} which was founded in 1963 to unite the TSOs of Denmark (Energinet), Finland (Fingrid), Iceland (Landsnet), Norway (Statnett) and Sweden (Svenska kraftnät) \cite{Nordel}. Since 2009, all Nordic TSOs are instead members of the \textit{European Network of Transmission System Operators}, ENTSO-E. The scope of this article is the high-voltage system, on transmission (400–220 kV) and subtransmission (150–110 kV) level. The latter voltage level is normally operated by TSOs, but this is not the case in Sweden -- hence the three major Swedish subtransmission DSOs operating on a 130 kV level (E.ON, Vattenfall and Ellevio) are also involved as stakeholders.

\begin{figure}[htbp]
    \centering
    \includegraphics[width=0.38\textwidth]{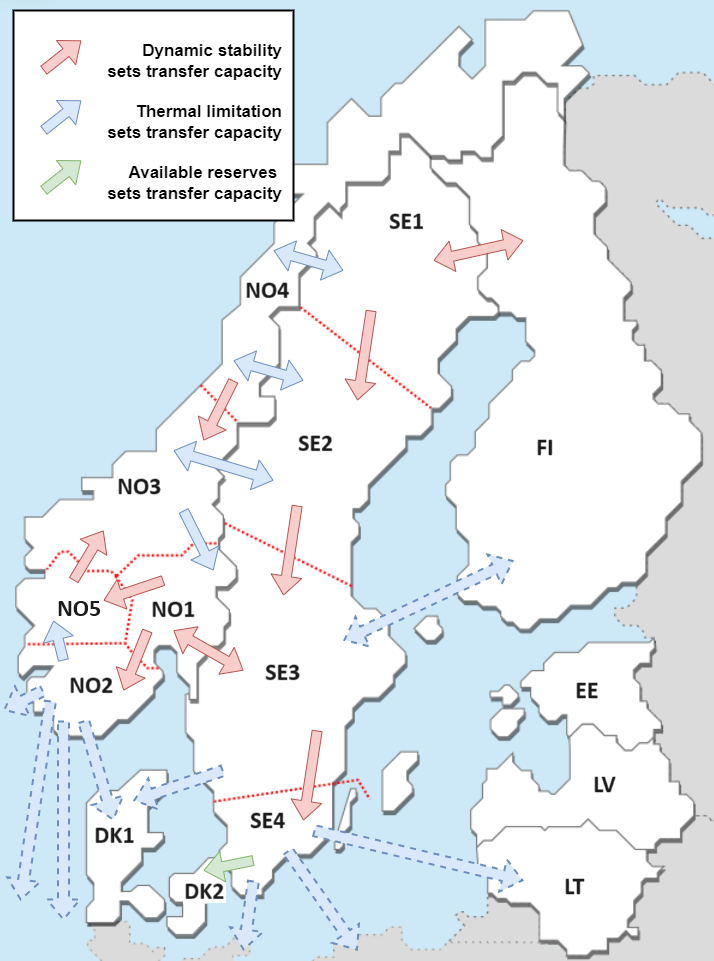}
    \caption{Map of bidding zones in the Nordics and Baltics, and the phenomena that typically limit transfer capacities on Swedish, Norwegian and Finnish BZBs. Arrows with dashed outlines are HVDC links.}
    \label{fig:ENTSOEconstraints}
\end{figure}

Fig.~\ref{fig:ENTSOEconstraints} shows the electricity bidding zones in the Nordics and Baltics. The bidding zone borders (BZBs) between these represent structural constraints in the grid, i.e. bottlenecks. Sweden and Finland have a relatively strong north-south axis compared to Norway, where the electrification was historically more decentralised and the transmission grid became nation-wide later \cite{NorwayElectricityHistory}.

Fig.~\ref{fig:energy_transfer_nordic} below gives an overview of where the main transmission corridors in the Nordic system are. It shows the gross energy transfer across the ten BZBs with the highest yearly energy transfers in the Nordic system, both internal and external, sorted by maximum value in the six-year period 2018-2023. The graph is based on hourly data from the ENTSO-E transparency platform \cite{ENTSOETransparency}. The internal Swedish borders SE2 $\rightarrow$ SE3 and SE3 $\rightarrow$ SE4 are the BZBs with the highest yearly energy transfers not only in the Nordics but in fact in the whole ENTSO-E region \cite{ENTSOETransparency}. Another important BZB is NO5 $\rightarrow$ NO1, transmitting power from the hydro heavy NO5 to the load heavy Oslo region in NO1.

\begin{figure}[htbp]
    \centering
    \includegraphics[width=0.48\textwidth]{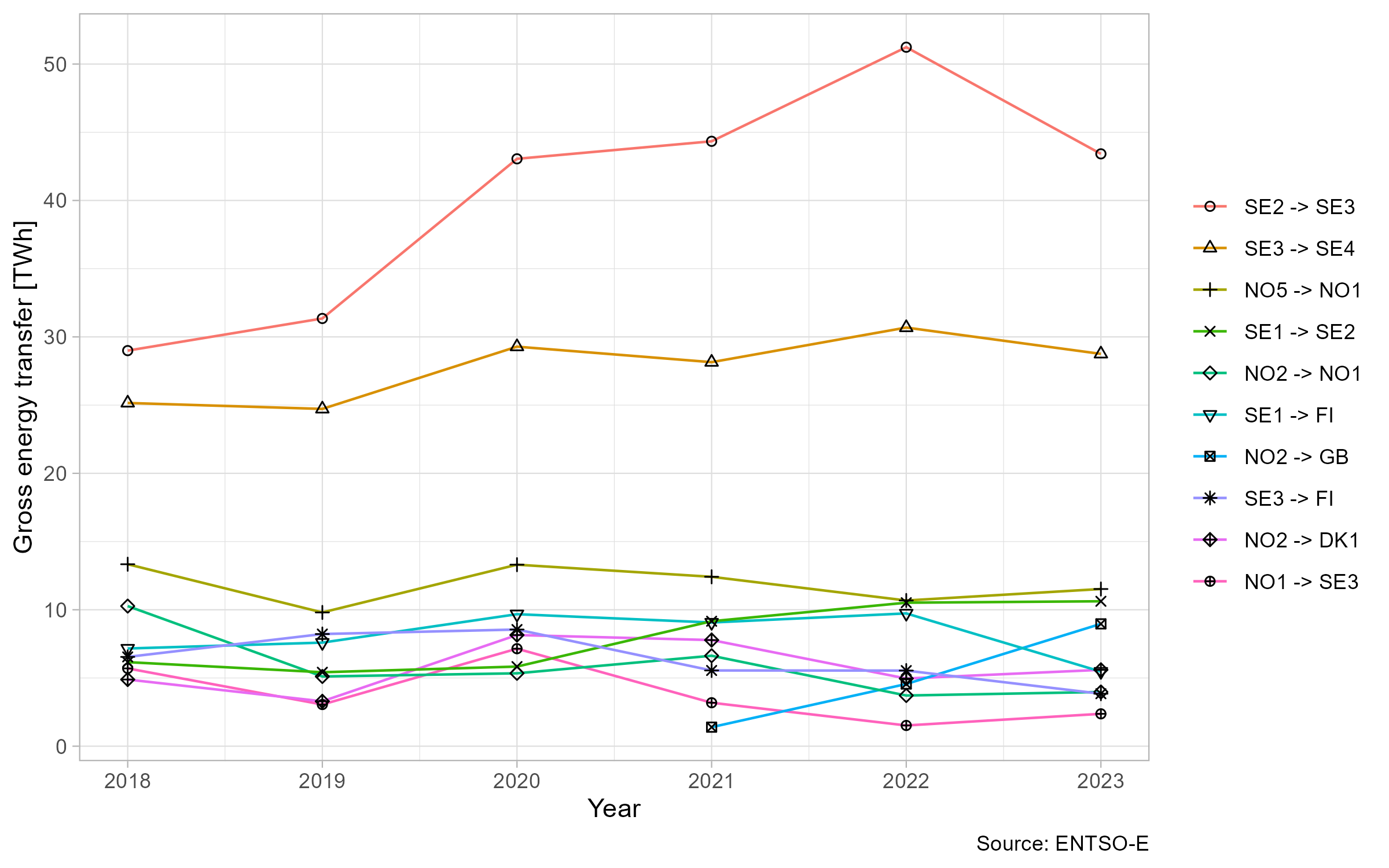}
    \caption{Yearly energy transfer across Nordic BZBs (top 10).  \cite{ENTSOETransparency}.}
    \label{fig:energy_transfer_nordic}
    \vspace{-0.4cm}
\end{figure}

\subsection{SIPS in relation to capacity allocation}
Already in the Nordic Grid Code from 2007 by \textit{Nordel}, it is stated that the Nordic power system is characterised by long transmission distances and relatively weak coupling between the subsystems. Weak coupling implies that it is not possible to use the full thermal capacity on some interconnections, rendering insufficient voltage support and/or insufficient damping as the limiting factor on the transfer capacity. If the limits are not respected, a single fault can result in voltage collapse (voltage instability) or generators losing synchronism (angle instability). The 2007 grid code specifically mentions SIPS as a way to increase the transmission capacity without building new lines when voltage or angle stability conditions set the capacity limit \cite{Nordel2007}. 

As seen in Fig.~\ref{fig:ENTSOEconstraints}, the Nordic power system is still largely limited by dynamic constraints. The information regarding the Swedish and Norwegian BZBs is gathered from the \text{ENTSO-E} Yearly Report About Critical Network Elements Limiting Offered Capacities \cite{SvkCNE2019}. The information regarding the Finnish BZBs is provided through discussion with Fingrid. Although the presentation of constraining phenomena in Fig.~\ref{fig:ENTSOEconstraints} is partial and crude, it highlights the need to include dynamic phenomena in capacity calculations. Also in the ACER document outlining the flow-based capacity calculation methodology for the Nordic region (2019), the need for dynamic allocation constraints on the SE2-SE3 border is emphasised. The transfer capacity on SE2-SE3, it states, can not be determined solely by its critical network elements (CNEs), but must be determined jointly through a combined dynamic constraint \cite{ACERflowbased}.

Regarding SIPS requirements in capacity calculations, according to the final amendment of the common coordinated capacity calculation methodology agreed upon by \text{ENTSO-E} and the Nordic TSOs, the TSOs must define RAs (including SIPS) to be taken into account in capacity calculation \cite{2020NordicManagement}.

\section{Method}
As described in sec. \ref{sec:aimofstudy}, the overarching goal of this report is to compare SIPS usage between different actors in the Nordic countries, focusing on, among other things, total deployment, targeted conditions and organisational philosophy. Additionally, we are interested in what has informed certain decisions regarding SIPS usage, such as views on risk assessment, perceived value and opportunities, and limitations related to grid structure. This motivated a literature review -- including articles and reports related to SIPS, regulations, agreements, etc. Additionally, a survey was sent out to a defined group of Nordic stakeholders and experts, and in-depth interviews were then conducted with them one by one.

The involved stakeholder group includes all five TSOs in the Nordic countries, three Swedish subtransmission DSOs (of which two responded), and two industry stakeholders, namely DNV and Hitachi Energy.

\section{Results and analysis}
In this section, the results from the literature, survey and interview study will be outlined. Due to the extensive amount of data collected, only some of the results are presented, and they are often analysed and compared as they are introduced. In Sec. \ref{sec:stocktaking}, results from the stocktaking of SIPS will be presented, both in quantitative and qualitative form. In Sec. \ref{sec:comparison}, a comparison between the different implementations and philosophies in the Nordics is made.

\subsection{Stocktaking of SIPS in the Nordics}
\label{sec:stocktaking}
\begin{figure}[htbp]
    \centering
    \includegraphics[width=0.48\textwidth]{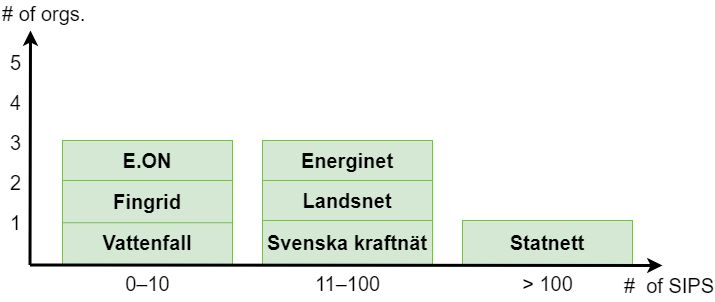}
    \caption{Histogram with estimated number of SIPS in own system.}
    \label{fig:NrOfSIPS}
    \vspace{-0.1cm}
\end{figure}

The estimated number of SIPS that each operator has implemented in their system is presented in Fig.~\ref{fig:NrOfSIPS}. As previously mentioned, the range and complexity of ``one SIPS'' can vary widely. Here, we have used the definition that all measurements and control actions used to mitigate the same critical system condition are considered to constitute the same SIPS. This can for example be overload of a line section or transmission corridor, under-voltage in a region or under-frequency in the whole synchronous area. All operators have counted e.g. their UFLS scheme only once. Due to difficulties in distinguishing between different SIPS, intervals rather than exact values are presented in Fig.~\ref{fig:NrOfSIPS}. 

Based on the survey and interviews, as well as publicly available information, it becomes clear that Statnett is the operator with the highest number of SIPS in the Nordics by a large margin. This can be confirmed by the results in Fig.~\ref{fig:NrOfSIPS}, where Statnett displays a number of SIPS in the hundreds. The majority of Statnett's SIPS are generation rejection, load shedding or grid reconfiguration schemes to prevent thermal overload on specific line sections or power transfer corridors. Svenska kraftnät (Svk) comes in second, with a growing number of generation rejection schemes (GRS) that enable connection of additional generation, mainly wind farms. These GRS are partly intended to mitigate overloads, but also ferroresonances and oscillations. Fingrid exhibits the least amount of SIPS in the Nordics, stating during their interview that they have almost no SIPS today and that the discussion on SIPS at Fingrid has only started. Landsnet, electrically isolated from other systems, is also an interesting case. In contrast to the other Nordic operators, they have embraced the WAMPAC approach, using phasor measurement units (PMUs) for both monitoring and control. 

In Tab. \ref{tab:condition_action}, the implemented SIPS in the Nordics are presented from a more qualitative viewpoint. Here, each combination of critical system condition (input) and mitigative action (output) is presented per operator. As seen in Tab. \ref{tab:condition_action}, Svk exhibits the largest variability in SIPS input-output combinations, with 17 unique combinations. The lack of voltage support in southern Sweden is for example visible in the implemented SIPS, with four different mitigative actions against abnormal voltage, and four against voltage instability. Svk is also the only operator with SIPS in place against small-signal angle instability, and with FACTS control as a mitigative action. Some combinations unite a majority of the operators, such as the UFLS scheme (``frequency instability-load shedding'') that all operators take part in, but also generation rejection and grid reconfiguration to mitigate component overload are common. Most combinations however are only ticked by one or two operators, indicating a diversity in solutions.

\begin{table*}[htbp!]
\renewcommand{\arraystretch}{1.2}
\caption{SIPS input-output combinations by operator \\ Actions are sorted by increasing cost, conditions by criticality level or range \\ 1 = Energinet, 2 = Fingrid, 3 = Landsnet, 4 = Statnett, 5 = Svenska kraftnät, 6 = E.ON, 7 = Vattenfall}
\label{tab:condition_action}
\begin{center}
\vspace{-0.3cm}
\begin{tabular}{|c|c|c|c|c|c|c|c|c|}
\hline
\textbf{Critical}&\multicolumn{7}{|c|}{\textbf{Mitigative action}} \\
\cline{2-8}
\textbf{system} & Grid & FACTS & Var rescheduling / & HVDC & Gen. rescheduling / & Generation & Load \\
\textbf{condition} & reconfiguration & control & Reactive power control & control & Active power control & rejection & shedding \\
\hline
Component overload& 2, 3, 4, 5, 7 & & & 1, 2 & 1, 3, 6 & 1, 4, 5, 6, 7 & 3, 4 \\
\hline
Abnormal voltage& & 5 & 5, 6, 7 & 5 & & & 3, 5 \\
\hline
Transient angle instability& 1, 3, 5 & & 1 & 1 & 1, 3, 5 & 5 & 3 \\
\hline
Small-signal angle instability& 5 & & & & 5 & & \\
\hline
Voltage instability& 1, 3, 5 & 5 & 1, 5 & 1, 2, 4, 5 & 1 & 6, 7 & 3 \\
\hline
Frequency instability& 1, 3 & & & 2, 4, 5 & 3 & & All \\
\hline
\end{tabular}
\label{tab1}
\end{center}
\vspace{-0.4cm}
\end{table*}

\subsection{Comparison of SIPS in the Nordics}
\label{sec:comparison}
In this section, the usage of SIPS in the Nordics is compared on the three different aspects mentioned in Sec. \ref{sec:aimofstudy}.
\subsubsection{Purpose (capacity vs. reliability)}
\label{sec:purpose}
As mentioned in the introduction, SIPS can be used both for increased transfer capacity, increased reliability, or a combination of the two. From the survey, connection of additional generation was also found to be an important objective, on the rise with more grid connection requests from distributed energy resources.

Regarding the purpose of SIPS, we identify both differences and shifts in attitude among the Nordic operators. Many operators anticipate increased use of automated system protection to enhance reliability in the future, see for example the system perspective analysis by Energinet from 2022 \cite{EnerginetSystemperspektiv}. Fingrid in turn recently raised the question in an internal discussion whether EU grid code allows SIPS also for increased transfer capacity, not only reliability.

\subsubsection{Implementation and maintenance}
\label{sec:implementation}
Both Svk and Fingrid are worried about the maintenance aspect of SIPS, and SIPS being left in the system that will either fail to operate or operate spuriously when the system is changing. Fingrid for example described an old SIPS in northern Finland which is incompatible with today's system and in the process of being dismantled. This serves as a reminder that renewing large infrastructures takes time, and that modern solutions might not always be compatible with old hardware and equipment. 

\subsubsection{Philosophy and risk assessment}
\label{sec:philosophy}
The philosophy regarding SIPS and their potential to contribute varies largely between the operators. Where Statnett has fully embraced and formalised the use of SIPS, at least local and event-based, Svk and Fingrid still express some hesitance. At Statnett, a relatively extensive use of SIPS is combined with frequent manual arming/disarming, changing the settings of their SIPS approximately 2000 times per year. This is admittedly becoming an issue, where Statnett are worried about increased complexity of their SIPS \cite{Ostli2023}. The philosophy at Statnett however is that SIPS are needed, and that the problems encountered due to the manual procedures should be solved through automation.

Several other actors are also worried about the increased complexity that SIPS will introduce. Unwanted interaction is seen as a potential risk, and with a quickly changing grid, there is a risk that existing SIPS will not work due to changed prerequisites. E.ON, who is in the process of updating their technical guidelines on SIPS, do for example not want SIPS in the meshed part of their grid for the time being, due to the risk of potential race conditions and unknown interactions.

Regarding PMU-based WAMPAC schemes, we perceive all Nordic TSOs except Landsnet to be more or less hesitant. In \cite{NGDP2023} for example, SIPS and particularly WAMPAC are depicted as risky and limited solutions. This attitude is in stark contrast to that given by Hitachi, DNV and Landsnet, who all stated that they think PMU data should be used more, also as input for control actions. Landsnet have collected PMU data for around 20 years, used it as control input for around 10 years and see great improvements compared to the old SCADA system. Both Hitachi and DNV mention reliable communication and cybersecurity as the main obstacles.

Few respondents perform any probabilistic risk assessments before implementing new SIPS. Only Energinet stated that they make a cost-benefit analysis before implementation. Both literature \cite{Liu2018} and DNV suggest that failure to operate when required is generally worse than spurious operation. This favours dependable, redundant and robust logic. When the system is reliant on SIPS, it has to work as intended, with at least the same reliability as classical fault clearing equipment.

\section{Conclusions}
\label{sec:conclusions}
Considering the large impact of SIPS on the power system as a whole, SIPS is rarely just a regional or national issue. All existing and future SIPS in the Nordic system should thus as a first step be integrated in the common grid model that is being developed by the coordinated capacity calculator, Nordic RCC. Going forward, the Nordic TSOs would benefit from increased cooperation and coordination of their SIPS. There is also plenty to learn from global experience.

We perceive most TSOs in the Nordic system to be more or less positive towards local, event-based SIPS. Statnett has gone the furthest, with hundreds of SIPS to increase capacity while avoiding overload. Svk is also accelerating their implementation, mainly driven by GRS to connect wind. Connection of additional generation has been mentioned as an important purpose also by other operators.

Moreover, we conclude that Statnett, and potentially also others, are struggling with the complexity related to having a large amount of event-based SIPS in their system. Several organisations also mention that the maintenance of SIPS can become an issue when the complexity increases, especially if documentation is missing. This could exacerbate the perceived problem of interactions, as there might be hesitance to decommission poorly understood SIPS in case they fill a vital function. This highlights the importance of correct and updated documentation, and performing recurrent testing and validation of all components in the system. The risk of maloperation could be more thoroughly analysed as many TSOs do not make any probabilistic risk assessment before implementing new SIPS.

When it comes to response-based, wide-area SIPS however, it is clear that most Nordic TSOs are still cautious. Complexity, vulnerability and cybersecurity are mentioned as obstacles. Other stakeholders, such as Landsnet and DNV, question whether distributed, wide-area SIPS are actually more complex or vulnerable than today's SCADA system. An advantage of WAMPAC vs. the traditional event-based approach is that the SIPS can become more intuitive. The reason for this is that the system response to both the contingency and the corresponding control action can be measured directly, thus eliminating the need for large amounts of distributed logic \cite{Sattinger2023}. It is worth noting in this context that the Nordic system has a large amount of HVDC links and other IBRs, with great controllability that could be exploited further.

To conclude, the electricity sector is changing rapidly, and definitions and practices need to be re-evaluated and updated. Additionally, we perceive that there is room for extension, coordination and optimisation of SIPS in the Nordics.

\section*{Acknowledgment}
We would like to thank all SPS4SE reference group members involved in the survey and interview study, for putting in both their time and expertise to realise this work.

\bibliographystyle{ieeetr}
\bibliography{SPS4SEref}

\begin{thebibliography}{10}

\bibitem{IEAelectrification2024}
{IEA}, ``Electrification.'' \url{https://www.iea.org/energy-system/electricity/electrification}, 2024.
\newblock [Accessed 2024-04-23].

\bibitem{gridliftoff2024}
{L. White et al.}, ``Pathways to commercial liftoff: Innovative grid deployment,'' tech. rep., U.S. Department of Energy, April 2024.
\newblock \url{https://liftoff.energy.gov/innovative-grid-deployment/}. [Accessed 2024-04-23].

\bibitem{Stankovic2022}
S.~Stanković, E.~Hillberg, and S.~Ackeby, ``{System Integrity Protection Schemes: Naming Conventions and the Need for Standardization},'' {\em Energies}, vol.~15, no.~11, 2022.

\bibitem{Acker2001}
V.~Acker, P.~Cholley, P.~Crossley, C.~Taylor, and C.~Vournas, ``{System Protection Schemes in Power Networks},'' {\em Cigré 38.02.19}, 6 2001.

\bibitem{IEEEguide2020}
{IEEE Power System Relaying and Control Committee}, ``{IEEE Guide for Engineering, Implementation, and Management of System Integrity Protection Schemes},'' {\em IEEE Std C37.250-2020}, 2020.

\bibitem{CigreControl2018}
{Y. Fang et al.}, {\em A proposed framework for coordinated power system stability control}.
\newblock Cigré, 2018.
\newblock Reference: 742 - 2018.

\bibitem{EU_CACM}
{Regulation 2015/1222}, ``{\textit{Commission Regulation (EU) 2015/1222 of 24 July 2015 establishing a guideline on capacity allocation and congestion management (CACM)}}.'' \url{http://data.europa.eu/eli/reg/2015/1222/2021-03-15}, 2015.
\newblock [Accessed 12-04-2024].

\bibitem{WAMPAC2019}
S.~Skok and I.~Ivankovic, ``{System Integrity Protection Schemes for Future Power Transmission System Using Synchrophasors},'' in {\em 2019 International Conference on Smart Grid Synchronized Measurements and Analytics (SGSMA)}, pp.~1--7, 2019.

\bibitem{IeaEnergyStatistics}
{IEA}, ``{E}nergy {S}tatistics {D}ata {B}rowser.'' \url{https://www.iea.org/data-and-statistics/data-tools/energy-statistics-data-browser}, 2023.

\bibitem{worldbank_capitaconsumption}
{World Bank}, ``{Electric power consumption (kWh per capita) - Sweden, Norway, Finland, Denmark, European Union}.'' \url{https://data.worldbank.org/indicator/EG.USE.ELEC.KH.PC?contextual=default&end=2014&locations=SE-NO-FI-DK-EU&start=1960&view=chart}, 2014.

\bibitem{Nordel}
{ENTSO-E}, ``{Former Associations -- NORDEL}.'' \url{https://docstore.entsoe.eu/news-events/former-associations/nordel/Pages/default.aspx}, 2018.
\newblock [Accessed 12-04-2024].

\bibitem{NorwayElectricityHistory}
{Norges vassdrags- og energidirektorat}, ``{Overview of Norway's Electricity History}.'' \url{https://publikasjoner.nve.no/rapport/2017/rapport2017_15.pdf}, 2016.
\newblock [Accessed 19-03-2024].

\bibitem{ENTSOETransparency}
{ENTSO-E}, ``{E}{N}{T}{S}{O}-{E} {T}ransparency {P}latform.'' \url{https://transparency.entsoe.eu/}.
\newblock [Accessed 19-03-2024].

\bibitem{Nordel2007}
Nordel, ``Nordic grid code 2007.'' \url{https://eepublicdownloads.entsoe.eu/clean-documents/pre2015/publications/nordic/planning/070115_entsoe_nordic_NordicGridCode.pdf}, 2007.
\newblock [Accessed 19-03-2024].

\bibitem{SvkCNE2019}
ENTSO-E, ``{Critical Network Elements -- Yearly Report About Critical Network Elements Limiting Offered Capacities [11.4]}.'' \url{https://transparency.entsoe.eu/transmission-domain/r2/yearlyReportCriticalElement/show}, 2024.
\newblock [Accessed 19-03-2024].

\bibitem{ACERflowbased}
{ACER}, ``{Long-term capacity calculation methodology of the Nordic capacity calculation region}.'' \url{https://nordic-rcc.net/wp-content/uploads/2024/01/FCA-Nordic-CCR-LT-CCM-2019-1.pdf}, 2019.

\bibitem{2020NordicManagement}
{ENTSO-E}, ``{\textit{Nordic Capacity Calculation Region capacity calculation methodology in accordance with Article 20(2) of Commission Regulation (EU) 2015/1222 of 24 July 2015 establishing a guideline on capacity allocation and congestion management}},'' tech. rep., 2020.

\bibitem{EnerginetSystemperspektiv}
{Energinet}, ``{Systemperspektivanalyse 2022 -- Udviklingsveje Mod Fremtidens Robuste Energisystem}.'' \url{https://energinet.dk/media/dnel5zlu/systemperspektivanalyse-2022.pdf?la=da&hash=68082DA51505444089EC98CD0AA8F68171E4EA46}, 2022.

\bibitem{Ostli2023}
``{Statnett om systemdriften: – Hjernen koker over}.'' \url{https://www.europower.no/nett/statnett-om-systemdriften-hjernen-koker-over/2-1-1539001}, 2023.
\newblock [Accessed 19-03-2024].

\bibitem{NGDP2023}
{Energinet, Fingrid, Statnett, Svenska kraftnät}, ``{Nordic Grid Development Perspective 2023}.'' \url{https://www.svk.se/siteassets/om-oss/rapporter/2023/svk_ngpd2023.pdf}, 2023.
\newblock [Accessed 20-03-2024].

\bibitem{Liu2018}
N.~Liu and P.~Crossley, ``Assessing the risk of implementing system integrity protection schemes in a power system with significant wind integration,'' {\em IEEE Transactions on Power Delivery}, vol.~33, pp.~810--820, 4 2018.

\bibitem{Sattinger2023}
{W. Sattinger et al.}, ``{Wide Area Monitoring Protection and Control Systems - Decision Support for System Operators Power system operation and control},'' 2023.
\newblock Cigré, C2 - CIGRE Technical Brochure 917.

\end{thebibliography}

\end{document}